\documentclass[useAMS,usenatbib]{mn2e}

\usepackage{amsmath}
\usepackage{comment}
\usepackage{graphicx}
\usepackage{rotating}
\usepackage{color}
\usepackage{aas_macros}
\usepackage{amsfonts}
\newcommand{\beq}{\begin{equation}}
\newcommand{\enq}{\end{equation}}
\newcommand{\beqa}{\begin{eqnarray}}
\newcommand{\enqa}{\end{eqnarray}}
\newcommand{\beit}{\begin{itemize}}
\newcommand{\enit}{\end{itemize}}
\newcommand{\bem}{\begin{pmatrix}}
\newcommand{\enm}{\end{pmatrix}}

\newcommand{\veck}{\mathbf{k} }

\newcommand{\lat}{\left\langle}
\newcommand{\rat}{\right\rangle}
\newcommand{\av}[1]{\lat #1 \rat}

\newcommand{\lp}{\left (}
\newcommand{\rp}{\right )}

\renewcommand{\max}{\mathrm{max}}
\renewcommand{\min}{\mathrm{min}}

\newcommand{\bes}{\begin{sideways}}
\newcommand{\ees}{\end{sideways}}

\title[Fisher matrix for nonlinear clustering]{On the information content of  the matter power spectrum}
\author[Carron, Wolk and Szapudi]{J. Carron$^{1,2}$\thanks{E-mail:
j.carron@sussex.ac.uk}, M. Wolk$^1$ and I. Szapudi$^1$  \\
$^1$Institute for Astronomy, University of Hawaii, 2680 Woodlawn Drive, Honolulu, HI, 96822, U.S.A.\\
$^{2}$Department of Physics and Astronomy, University of Sussex, Brighton BN1 9QH, U.K}
\begin{document}

\date{\today}

\pagerange{\pageref{firstpage}--\pageref{lastpage}} \pubyear{2014}

\maketitle

\label{firstpage}

\begin{abstract}
We discuss an analytical  approximation for the matter power spectrum covariance matrix and its inverse on  translinear scales, $k \sim 0.1h - 0.8h/\textrm{Mpc}$ at $z = 0$. We proceed to give an analytical expression for the Fisher information matrix of the nonlinear density field spectrum, and derive
implications for its cosmological information content. We find that the spectrum information is characterized by a pair of upper bounds, 'plateaux',  caused by the trispectrum, and a 'knee' in the presence of white noise. The effective number of Fourier modes, normally growing as a power law, is bounded from above by these plateaux, explaining naturally earlier findings from $N$-body simulations. These plateaux limit best possible measurements of the nonlinear power at the percent level in a $h^{-3}\textrm{Gpc}^3$ volume; the extraction of model parameters from the spectrum is limited explicitly by their degeneracy to the nonlinear amplitude. The value of the first, super-survey (SS) plateau depends on the characteristic survey volume and the large scale power; the second, intra-survey (IS)  plateau is set by the small scale power. While both have simple interpretations within the hierarchical \textit{Ansatz}, the SS plateau can be predicted and generalized to still smaller scales within Takada and Hu's spectrum response formalism. Finally, the noise knee  is naturally set by the density of tracers.
\end{abstract}


\newcommand{\Cov}{\textrm{Cov}}
\newcommand{\SSM}{\textrm{SSM}}
\newcommand{\Trispectrum}{\textrm{Tris.}}
\newcommand{\assm}{\sigma_{\textrm{SS}}}
\newcommand{\ahsv}{\sigma_{\textrm{IS}}}
\newcommand{\atot}{\sigma_{\min}}

\section{Introduction}

The current multi-faceted effort of the cosmology community to characterize cosmic acceleration by mapping the low-redshift Universe \citep{WeinbergEtal2013} fundamentally relies on a robust understanding of the statistics of the evolved matter density field and its tracers.  In the context of weak-lensing, baryon acoustic oscillations (BAO) or more generally large scale structure galaxy surveys, the key statistic is the power spectrum of the matter and galaxy field. In particular,
an accurate quantitative and qualitative understanding of the cosmic covariance of the spectrum  beyond Gaussian statistics \citep{FeldmanEtal1994,VogeleySzalay1996} is necessary for a  successful completion of this program.
A surprising recent insight is that the covariance can have a substantial contribution due to the coupling of small scales to modes of size comparable to or larger than the survey size, dubbed 'beat coupling' in \citep{HamiltonEtal2006}. This effect is not expected from early analytical work \citep{MeiksinWhite1999,ScoccimarroEtal1999}, and not captured correctly by most traditional ensemble average estimates of the covariance from simulations \citep[see the discussion in][]{HamiltonEtal2006}.
Furthermore, it is established that the non-Gaussian contributions generically create saturation in the information content of the spectrum on mildly nonlinear scales for amplitude-like parameters \citep{ScoccimarroEtal1999,RimesHamilton2005,RimesHamilton2006,NeyrinckEtal2006,NeyrinckEtal2007, TakahashiEtal2009}. See also \cite{LeePen2008}.  This stands in sharp contrast to naive expectations based on Fourier modes counting, and has potentially stark consequences for the available information from galaxy surveys, or indeed, for the effective volume accessible to us for constraining Dark Energy and other ingredients of the cosmological model. Moreover, these nonlinear effects are large enough to affect the covariance matrix even on the BAO scale \citep{TakahashiEtal2009,NganEtal2012}. Most traditional forecasting techniques based on the Fisher information content of Gaussian Fourier modes likely over-predict somewhat the scientific return of cosmological data sets.
\newline
\indent
A simple analytical approximation has been fit to the matter power spectrum covariance matrix
from $N$-body simulations \citep{Neyrinck2011,MohammedSeljak2014}. The principal aim of this paper is to discuss and exploit this approximation that captures the non-Gaussian effects on translinear scales $k \sim 0.1h - 0.8h/\textrm{Mpc}$ at $z = 0$, and for the first time, present its analytical inverse and obtain new insights into the information content of mildly non-linear scales. In particular, we obtain a simple formula for the Fisher information content of the spectrum down to translinear scales, extending the standard expression of \cite{Tegmark1997b} valid only for Gaussian fields. In the end, our analytical estimations explain in the simplest and most natural way the behavior of the information content of the spectrum on cosmological parameters.
\newline
\indent
The paper is built as follows. Section \ref{covmatrix} introduces our model for the covariance matrix and discusses its ingredients in light of the hierarchical model. Section \ref{info} derives  the Fisher information matrix of nonlinear clustering and explores the consequences. We conclude in section \ref{conclusion} with a discussion of our findings as well as of some aspects and extensions of the covariance matrix model.
\section{Approximate covariance matrix} \label{covmatrix}
We first introduce the key concept, an approximate form of the matter power spectrum covariance in the mildly non-linear regime and then discuss its properties.
\newline
\indent
Assuming a finite survey volume $V$ the standard shell-averaging estimator $\hat P(k)$ for the power spectrum is,
\beq
\hat P(k) = \frac{1}{V N_k} \sum_{\veck'} \left | \delta(\veck')\right|^2.
\enq
The sum runs over all of the $N_k$ Fourier modes assigned to the  $k$-shell corresponding to a power spectrum bin.
Let further the matrix $\Cov_{ij}$ be the covariance matrix of the estimator, 
\beq
\Cov_{ij} = \av{\hat P(k_i) \hat P(k_j)} - \av{ \hat P(k_i)}\av{ \hat P(k_j)}.
\enq
The working hypothesis of this paper is the following functional form for the covariance matrix
\beq \label{Cov}
\Cov_{ij} =  \delta_{ij} \frac{2 \lp P(k_i)+\frac 1 {\bar n}\rp^2}{N_{k_i}} + \atot^2P(k_i)P(k_j).
\enq
The first term is diagonal  corresponding to the Gaussian covariance. The second term approximates the shell-averaged trispectrum of the fluctuation field. As will be shown later the parameter $\atot^2$ has a natural interpretation as a minimum achievable  variance, hence its name. Given the complexity of the matter trispectrum, this approximation might at first sight appear to be a blunt hypothesis. It's justification lies entirely in fits to $N$-body simulations, used first by \cite{Neyrinck2011} and rediscovered by \cite{MohammedSeljak2014}. The latter authors quote an accuracy of $10-20\%$ on all matrix elements which is noteworthy for such a simple analytical form and certainly enough for our analysis focusing on generic properties of the power spectrum information.
\newline
\indent We will decompose  the second term further into two pieces as follows
\beq \label{coeff}
\atot^2 = \assm^2 + \ahsv^2.
\enq
The first term captures  the impact of large wavelength 'super-surveys' (henceforth SS) modes on the small scale covariance, and the second corresponds to small scale-small scale couplings, or the 'intra-survey' (IS) modes contribution to the trispectrum. We turn to a discussion of this intuitively non-trivial Ansatz next. 
\subsection{Gaussian covariance}
The Gaussian covariance simply counts the number of independent Gaussian variables in the initial Gaussian field. It depends the number of modes $N_k$ for which we will use a standard approximation: the number of modes is the volume of the shell used for the bin averaging, divided by the volume element between two discrete modes. In our Fourier conventions the latter is $(2\pi)^3/V$. It results 
\beq \label{Nk}
N_k \approx  V  \frac {4\pi k^2dk}{(2\pi)^3} 
\enq
where $dk$ is the width of the bin associated to $k$.
\newline
\indent
In order to discriminate the impact of discreteness effects typical of galaxy surveys data to cosmic variance, white noise is incorporated in a simple way in \eqref{Cov} through a term $1/ \bar n $, that enters the Gaussian covariance but neither the shell averaged trispectrum nor through additional terms. This prescription is exact in the handy and insightful case of perfectly Gaussian noise, itself uncorrelated to the $\delta$ field, with spectrum $1/\bar n$. Indeed, connected point functions such as the spectrum and trispectrum of independent fields always just add up, from which Eq. \eqref{Cov} directly follows. More realistic is the case of Poisson noise, in which case this prescription captures the leading effects. Poisson sampling generates additional covariance \cite[e.g]{MeiksinWhite1999, Smith2009} which are however all formally suppressed with respect to the leading correction implemented in \eqref{Cov}. Anticipating somewhat our results, the importance of this leading discreteness correction is very minor in comparison to that of cosmic variance for the purposes of this paper. We will therefore not study here the impact of additional discreteness corrections, nor other issues such as galaxy biasing.
\subsection{SS and IS covariance}
We postulated the form given by Eq. \eqref{Cov} for the covariance matrix. How to estimate roughly $\atot^2$ ? We develop some approximations next. Let us multiply the noise free Eq. \eqref{Cov} on both sides by $(N_{k_i}/V)\cdot (N_{k_j}/V)$. We then sum over all modes. The rightmost terms combine to $\atot^2 \sigma^4 $, where $ \sigma^2 = \av{\delta^2}$ is the variance of the field at the resolution considered.  The other terms form the two-point cumulants  averaged  over the survey volume, giving
\beq
\label{ctot}
\atot^2 = \frac{1}{\sigma^4} \int_V \frac{d^3x}{V} \int \frac{d^3y}{V} \av{\delta^2(x)\delta^2(y)}_{\textrm{c}}. 
\enq
Within the hierarchical \textit{Ansatz} (HA) \citep{Peebles1980,Fry1984b} or standard perturbation theory (SPT) in the weakly non-linear regime \citep{Bernardeau1996}, the leading contribution to the cumulant is
\beq
\av{\delta^2(x)\delta^2(y)}_c \sim 4R_a\sigma^4\xi(x-y). 
\enq
The proportionality constant $4 R_a$ is the amplitude of the four snake diagrams involved \citep{Fry1984b}. They correspond to $4\tilde Q_{22} = C_{22}$ in the notation of \cite{SzapudiSzalay1997} or \cite{Bernardeau1996} respectively.  Plugging this contribution into Eq. \eqref{ctot} gives a term proportional to the integral of the two-point function over the survey volume. This integral is nothing else than the variance of the background Fourier mode $\delta_b = \int_V dx \delta(x) /V$,
\beq
\begin{split}
& \int_V \frac{d^3x}V\int_V \frac{d^3y}V \xi(x-y) = \av{\delta_b^2} = : \sigma^2_V \\
 & = \int \frac {d^3k}{(2\pi)^3} P(k)W^2(\veck). \end{split}
\enq
In the last equation, $W(\veck)$ is the Fourier transform of the window function $W(x)$ of the survey volume, equal to $1/V$ if $x$ is in the volume, $0$ if not. We define this contribution proportional to $\sigma^2_V$ the SS coefficient $\assm^2$. Thus, in the HA \citep[c.f.,][]{SzapudiColombi1996}
\beq \label{cssm}
\assm^2  = \sigma^2_V \:4 R_a \approx \frac{ P(k_\min)} V 4R_a.
\enq
We have used there that the weight function $W^2(\veck)$ probes the power on the largest scales available in the survey volume, and has total weight $\int d^3k/(2\pi)^3 W^2(\veck) = 1/V$.
\newline
The remaining contributions to $\atot^2$ are assigned to an IS coefficient $\ahsv^2$. Within the HA the next contribution to the cumulant is of the form
\beq
 \sigma^2\xi^2(x-y) (4R_a + 4R_b). 
\enq
Plugging into Eq. \eqref{ctot} leads to
\beq
\ahsv^2 \sim  \lp \int_V \frac{d^3x}{V}  \int_V \frac{d^3y}{V} \frac{\xi^2(x-y)}{\sigma^2}  \rp(4R_a + 4R_b) .
\enq
In contrast to $\assm$, $\ahsv$ probes the smallest scales. This can be best seen after Fourier transformation.
From the same argument as above for the window function holds
\beq
\begin{split}
& \int_V \frac{d^3x}{V}  \int_V \frac{d^3y}{V} \xi^2(x-y)\\
 & = \int \frac{d^3k}{(2\pi)^3}P(k) \int \frac{d^3k'}{(2\pi)^3}P(k')W^2(\veck - \veck') \\
 & \approx  \frac 1 V \int \frac{d^3k}{(2\pi)^3}P^2(k) 
 \end{split}
\enq
On the other hand, in terms of the dimensionless spectrum $\Delta(k) = k^3P(k)/2\pi^2$ we get
\beq
\frac 1 {\sigma^2} \frac 1 V \int \frac{d^3k}{(2\pi)^3}P^2(k)  = \frac 1 V \frac{\int d\ln k \: \Delta(k) P(k)}{\int d\ln k \:\Delta(k)} \approx \frac{ P(k_\max)}{V}, 
\enq
where we used that the weight of $P(k)$ in that equation is peaked towards small scales in a $\Lambda$CDM universe, and the total weight is unity. Thus,
\beq \label{ahsv}
\ahsv^2 \approx  \frac{ P(k_\max)}{V}(4R_a + 4R_b).
\enq
Comparing to Eq. \eqref{cssm} there is an interesting symmetry between large and small scales.

\subsection{Local versus global density}
We have not discussed so far whether the fluctuation $\delta$ is defined with respect to the 'local', observed, density as in the case of galaxy surveys, or with respect to the global density, as appropriate, e.g., for weak lensing surveys. This can make a difference in what follows. The local density fluctuation is the global fluctuation divided by the background mode
\beq
\delta(x)   \textrm{   (local)} = \frac{\delta(x) \textrm{  (global)  }}{1 + \delta_b},
\enq
and the denominator gives a further contribution to the covariance. This contribution can be calculated from PT following \cite{DePutterEtal2012}, or from the modified response of the spectrum to a background mode \citep{TakadaHu2013} (hereafter TH13). This contribution is proportional to $\sigma^2_V$ and thus can be absorbed entirely within $\assm^2$, as discussed below.
\newline
\newline
\indent
Let us summarize our findings so far and estimate the size of these contributions. The covariance matrix of the power spectrum inside a survey volume is well characterized throughout the translinear scales by two dimensionless coefficients capturing large and small scale power
\beq
\assm^2 \sim \frac {P(k_\min)}  V \textrm{   and  } \ahsv^2 \sim\frac{ P(k_\max)} V 
\enq
up to factors that are related to the hierarchical amplitudes. Our SS contribution coincide with an already fairly well understood contribution in the literature, that can also be derived from PT \citep{HamiltonEtal2006} or the response formalism \citep[TH13]{LiEtal2014}. From $4R_a = C_{22} = (68/21)^2$ we get
\beq
\assm^{2} \sim  \begin{cases} & \sigma^2_V \lp \frac{68}{21}\rp^2 \textrm{     (global) } \\
                                                 &  \sigma^2_V \lp \frac{26}{21}\rp^2 \textrm{     (local) }. \end{cases}
\enq
The local case is obtained by subtracting $2$ to to the original root coefficient, here $68/21$ \citep[TH13]{DePutterEtal2012}. A typical value for $\sigma^2_V$ is $1.5\cdot10^{-5}$ for a spherical volume of $h^{-3}(\textrm{Gpc}^3$ (TH13, Fig. 1). The IS contribution is not as well understood, but we can estimate it from the arguments above. The power on translinear scales is down by one to two orders of magnitude from the power close to the peak of the $\Lambda$CDM spectrum. The hierarchical factors are only a factor of two or so larger for the IS coefficient. This shows that the SS contribution dominates in the global case. On the other hand, the reduction of the SS coefficient by one order of magnitude in the local case brings it down to the level of the IS effects. All this is perfectly consistent with what is currently known on the SS contribution to the covariance from simulations, as summarized for instance by \cite{DePutterEtal2012} or \cite{LiEtal2014}.

\section{Information within nonlinear clustering} \label{info}
Understanding information in the power spectrum requires inversion of its covariance matrix.
Now, why should this form of the covariance matrix be particularly useful ? Because Eq. \eqref{Cov} is the very prototype of a perturbation to a matrix that allows analytical inversion, through the Shermann-Morrison (SM) formula \citep[e.g.]{PressEtal07}: for an invertible matrix $M$ and vectors $u,v$ holds
\beq \label{SM}
(M + uv^T)^{-1} = M^{-1} - \frac{M^{-1}uv^TM^{-1}}{1 + v^TM^{-1}u}.
\enq
Defining  $u_i = v_i = \atot P(k_i)$,
and $M$ the diagonal Gaussian covariance, the RHS is the inverse of the covariance matrix \eqref{Cov}. The contraction $v^TM^{-1}u$ has a simple interpretation which we elaborate next.
\newline
\newline
\indent 
We introduce for notational convenience a $z = 0$ nonlinear spectrum amplitude parameter $\ln A_0$, through $\partial_{\ln A_0} P(k_i) = P(k_i)$, and thus $\partial_{\ln A_0 } \ln P(k_i) = 1$.  Explicitly, a formal definition of that parameter is $P(k,A_0) = A_0P(k)$, with the understanding that we work at the fiducial value $A_0 = 1$. The parameter $A_0$  can be identified with $\sigma^2_8$, or the initial amplitude in the linear regime, but is distinct on translinear scales considered here. The Fisher matrix element on this parameter is the squared cumulative signal to noise \citep{TakahashiEtal2009}, traditionally studied as a function of the resolution $k_\max$, that measures how well the spectrum itself can be measured. As we show below, the form of the covariance matrix causes this parameter to play a special role. Let the Fisher matrix of the spectrum be
\beq \label{SNeq}
F_{\alpha \beta} = \sum_{k_i,k_j \le k_\max} \frac{\partial P(k_i)}{\partial \alpha} \Cov^{-1}_{ij}\frac{\partial P(k_j)}{\partial \beta}, 
\enq
then $(S/N)^2= F_{\ln A_0 \ln A_0}$. With this in place we can calculate the Fisher information content of the spectrum. Contracting the SM formula on the left with $\partial_\alpha P(k_i)$, and on the right with $\partial_\beta P(k_j)$, we get
\beq
\label{FNG}
\begin{split}
F_{\alpha \beta}& = F^G_{\alpha \beta} - \atot^2 \frac{F^G_{\alpha \ln A_0}F^G_{\ln A_0 \beta}}{1 + \atot^2 (S/N)_G^{2}}, \\
\end{split}
\enq
one of the key result of this paper. There the Gaussian Fisher matrix elements $F^G_{\alpha \beta}$ are given by an already familiar formula \citep{Tegmark1997b}
\beq
\begin{split}
\label{FNGa}
 F^G_{\alpha \beta} =  \int d \ln k\: \omega(k)  \frac{\partial  \ln P(k)}{\partial \alpha}\frac{\partial  \ln P(k)}{\partial \beta}.
\end{split}
\enq
with
\beq \label{wk}
w(k) = \frac V 2\frac{k^3}{2\pi^2}\lp  \frac{P(k)}{P(k) + \frac 1 {\bar n}}\rp^2.
\enq
In these equations discrete sums over the averaging shells $k_i$ have been replaced in a natural way by an integral, using the approximation \eqref{Nk} for the number of modes. The denominator of the Fisher matrix is the Gaussian signal to noise
\beq \label{SNG}
\lp S/N\rp^2_{G} = \int d \ln k \:w(k) = F^G_{\ln A_0 \ln A_0},
\enq
obtained by contracting the inverse Gaussian covariance with the derivatives $\partial_{\ln A_0}P(k_i) = P(k_i)$.
\newline
\indent
We see from Eqs. \eqref{FNG} and \eqref{FNGa} that our choice of covariance matrix combines the Gaussian covariance with a projection $P(k_i)P(k_j)$ on the nonlinear spectrum, making in fact implicitly a strong assumption on the fluctuations in power: non-Gaussian covariance only affects the amplitude of the spectrum; fluctuations that are orthogonal to the amplitude, in the precise sense of 
\beq
\int \frac{dk}{k} \:\omega(k) \:\delta \ln P(k) = 0
\enq 
probe the Gaussian part of the covariance exclusively.
\newline
\indent
Of particular interest is the non-Gaussian signal to noise of the spectrum. Setting $\alpha = \beta = \ln A$ in Eq. \ref{FNG} gives us
\beq \label{SNLN}
(S/N)^2 = \frac{(S/N)^2_G}{1 + \atot^2 (S/N)^2_G}.
\enq
We now find an essential difference between the linear and nonlinear $S/N$. No matter how fast $(S/N)^2_G$ grows (following $k^3_\max$ in the noise-free field), the RHS of this equation is bounded by $1/\atot^2$: there is only a finite amount of information to extract. The inverse root information $1/\sqrt{F_{\alpha\alpha}}$ bounds the parameters variance $\sigma(\alpha)$, therefore the interpretation of $\atot$ becomes clear: 
\beq
\frac{ \sigma(A_0)}{A_0} \geq \atot.
\enq
From Eq. \eqref{SNLN}, the bound is saturated as soon as $\atot^2  (S/N)^2_G \gg 1$. We can estimate a saturation
\newcommand{\ksat}{k_{\textrm{sat}}}
scale $k_\textrm{sat}$, after which there is no more independent information on the amplitude in the noise free spectrum. We identify the plateau value with the Gaussian $(S/N)^2$ at $\ksat$, yielding $\ksat^3  = 12\pi^2 / V\atot^2$.
$\ksat$ shows some dependence on volume and survey geometry through $\assm$. For the spherical $h^{-3}\textrm{Gpc}^3$ volume $\sigma^2_V \sim1.5\cdot 10^{-5}$ we get 
\beq
\begin{split}
&\ksat \sim 0.09 h\textrm{Mpc}^{-1} \quad \frac{\sigma(A_0)}{A_0} \sim 1.2\%   \quad   \textrm{  (global)  }  \\
&\ksat \sim 0.13 h\textrm{Mpc}^{-1} \quad \frac{\sigma(A_0)}{A_0} \sim 0.7\% \quad \textrm{  (local) }
\end{split}
\enq
In the local case we used for the IS coefficient $\ahsv \sim \assm \sim \sigma_V \sim 0.005$. How about the linear fluctuation amplitude $\sigma_8$ ? One expects the degeneracy between $\sigma^2_8$ and $A_0$ to be somewhat lifted beyond linear scales, so there is in principle no absolute lower limit but  only a pronounced reduction in information. Straightforward evaluation of Eq. \eqref{FNG} with nonlinear power evaluated according to HALOFIT \citep{SmithEtal2003,TakahashiEtal2012}  as implemented in the CAMB\footnote{http://camb.info} software  \citep{LewisEtal2000} gives ($k_\max= 0.8h/\textrm{Mpc}$)
\beq
\begin{split}
\frac{\sigma(\sigma_8)}{\sigma_8} \sim 0.25\% \textrm{  (global)     } 0.17\%   \quad   \textrm{  (local)  }.  \\
\end{split}
\enq
We can try and make a connection to the two papers fitting this form of covariance matrix. First, \cite{MohammedSeljak2014} quote a higher value of $0.0079$ for what we interpret as the $\ahsv$ coefficient  (see their Eqs. 34-36) from a fit to simulations by \cite{LiEtal2014}. This would shift slightly the values obtained in the local case, to $0.9\%$ compared to our $0.7\%$ for $A_0$, and $0.2\%$ instead of $0.17\%$ for $\sigma_8$. On the other hand comparison to \cite{Neyrinck2011} is less simple. There the evaluation of the covariance matrix is made from single realizations with the method of \cite{HamiltonEtal2006}. While there is no background mode variance $\sigma^2_V$ in single such realizations, the SS effect is nevertheless captured in that method, at least partially, due to the elaborated weighting scheme probing the fundamental mode. Following this line of reasoning, Neyrinck's parameter $\alpha = \av{b^2}$ (c.f., his Eq. ~1) should be best compared to our $\atot^2$ in the global case. We would predict $\atot^2 \sim 1.5\cdot10^{-4}$, not too far from his quoted values of $\alpha \sim 1.2-1.4 \cdot 10^{-4}$. 
\newline
\indent It is important to note that the presence of a plateau is independent of the fine details of the covariance matrix, provided one contribution to it is of the form $\propto P(k_i) P(k_j)$. In the SM formula we can pick $u$ and $v$ to contain only the SS part of $\atot$, assigning to $M$ all remaining covariance, concluding similarly  that the measurement of the nonlinear amplitude is limited by $\assm$, irrespective of the exact form of the IS trispectrum. Likewise, irrespective of the details of the impact of the super-survey modes on the covariance, the measurement is bounded by $\ahsv$. This amounts to the existence of two distinct plateaux, a SS plateau and an IS plateau.
\begin{center}
\begin{figure}
\includegraphics[width = 0.4\textwidth]{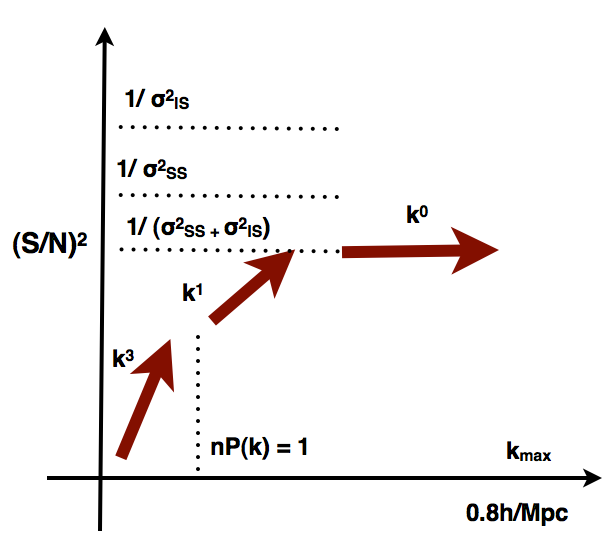}
\caption{\label{figsng} A schematic description of the behavior of the information content of the matter power spectrum in the mildly nonlinear regime, according to the spectrum Fisher matrix for the non-Gaussian field derived in this work, Eq. \eqref{FNG}. Shown is the squared signal to noise $(S/N)^2$, or effective number of Fourier modes of the spectrum. The curve transitions from power-law growth in the Gaussian regime to complete saturation. The plateau is due both to the effects of super-surveys (SS) modes and that of intra-survey (IS) modes. The former dominates for density fluctuations defined with respect to the global density. Both are of comparable value if the fluctuations are defined with respect to the local density.  The impact of white noise can only causes a shift in power-law growth exponent on these scales. The variance $\assm^2 + \ahsv^2$ is the minimal achievable error on the nonlinear power amplitude. The spectrum information content on other parameters is modulated by their degeneracies to the nonlinear amplitude according to Eq. \eqref{FNG}.}
\end{figure}
\end{center}
\indent How does all this compare to the impact of white noise on $(S/N)$? From representation \eqref{wk} we gather that
\beq
(S/N)^2_G  \propto \begin{cases} k^3 \quad &\bar n P(k) \gg 1 \\  k^{3 + 2n}  \quad &\bar n P(k) \ll 1, \end{cases}
\enq
where $n$ is the local slope of the spectrum.
We conclude that white noise shows in fact a milder impact on translinear scales : there the slope of the power spectrum is close to $-1$. This implies that $(S/N)^2_G$ still grows following $\propto k_\max$. Thus, no plateau but rather a knee, that occurs at $\bar n P(k)  \sim 1$, where the slope of $(S/N)^2$ changes from $3$ to $1$. On the other hand, deeper in the non-linear regime $k \sim h/\textrm{Mpc}$, where the spectrum slope is steeper, white noise causes indeed a plateau. For a galaxy number density of $\bar n \sim 3\cdot10^{-4}h^3\textrm{Mpc}^{-3}$  \citep{PercivalEtal2007} and a vanilla $\Lambda$CDM (unbiased) power spectrum the knee occurs roughly at $k \sim 0.15 h/\textrm{Mpc}$, which is comparable to $\ksat$, and is substantially pushed towards still smaller scales accounting for galaxy bias. Thus, it appears that shot noise should not be the dominant source of information loss on translinear scales. This conclusion may differ certainly for other types of white noises such as shape noise in weak-lensing. For sparse tracers there is also the theoretical possibility of a dominant white noise component on very large scales, on the other side of the $\Lambda$CDM spectrum peak.

\section{Discussion} \label{conclusion}
We proposed and explored the form \eqref{Cov} for the spectrum covariance matrix and its inverse. This simple form combines the Gaussian covariance with a projection on the nonlinear spectrum. It is remarkably successful at explaining the dynamics of the spectrum information observed in simulations, particularly the saturation of the effective number of Fourier modes $(S/N)^2$ as observed in simulations since the works of \cite{RimesHamilton2005,RimesHamilton2006}. Saturation is the most fundamental prediction of the model, with value $1/\atot^2 = 1/(\assm^2 + \ahsv^2)$. The nonlinear amplitude of the power spectrum cannot possibly be measured better than $\atot$, typically close to the percent level ($\sim 0.7-1.3\%$) in a $h^{-3}\textrm{Gpc}^3$ volume, depending on whether fluctuations are defined with respect to the local or global density. Degeneracies to the nonlinear amplitude decrease the information on other parameters according to Eq. \eqref{FNG}; for that same volume and in the absence of white noise the relative accuracy achievable on the linear root amplitude $\sigma_8$ is closer to $0.17-0.25\%$ (see also \cite{MohammedSeljak2014}). 
\newline
\indent
Eq. \eqref{FNG} presents the Fisher information matrix of the 3D matter spectrum on translinear scales, considering white noise, super-survey covariance (with an explicit dependence on survey geometry and volume) and intra-survey covariance. It comes at the same  implementation costs, very low by modern standards, as the traditional formula for Gaussian fields such as \cite{Tegmark1997b}. Thus, if applicable, it should naturally be preferred for parameter constraints forecasts beyond the linear regime. In contrast to the formula for Gaussian fields, it is by nature rather a conservative estimate if a reasonable choice of $\atot$, thus of nonlinear $S/N$ saturation scale, is performed. As weak-lensing data is more usually investigated in 2D, it is a natural direction for future work to investigate the pertinence of a similar form in that case. We note that this information content \eqref{FNG} is associated to traditional shell averaging estimators. It might be that some more sophisticated spectrum estimators can recapture more information, in analogy to the consideration of the local density in the local case. In general, a tentative reconstruction of the local density helps alleviating the SS plateau \citep{LiEtal2014b,CarronSzapudi2014b}.  In addition, the use of nonlinear transforms in the form of sufficient statistics further lower cosmic variance associated with the plateaux \citep[e.g.,][and references therein]{Neyrinck2011b,Neyrinck2011,Carron2012,CarronSzapudi2014,WolkEtal2014}.
\newline
\indent
This form of the covariance was suggested already by $N$-body simulations \citep{Neyrinck2011,MohammedSeljak2014}, and we did not venture into a full physical explanation of its origin. However, both coefficients can be associated to hierarchical amplitudes in simple ways. Our derivation leads in a very natural manner to a generically dominant SS contribution in the case of the global density, due to the conjunction of a comparatively strong power on large scales $\sigma^2_V$ and the strong coupling to small scales through the coefficient $4R_a \sim C_{22} \sim \tilde Q_{22} \sim 10$. It is interesting to note that lognormal field statistics, useful to simulate the information plateau and covariance matrices \citep[e.g.]{TakahashiEtal2014,CarronEtal2014}, underestimate the impact of non-Gaussian effects, $\assm^2/\sigma^2_V = 4$ instead of 10 in the global case (lognormal field statistics amount in this context to setting $R_a = R_b = 1$ on mildly nonlinear scales). In the local case, the lognormal field has in fact $\assm = 0$. Note that we did not consider smoothing effects in our analysis; as is known \citep[e.g.]{BernardeauKofman1995,CarronSzapudi2013} smoothing brings the matter density field closer to lognormal behavior on mildly nonlinear scales. 
\newline
\indent
\cite{Neyrinck2011} observed that this form of the covariance may be interpreted through bias fluctuations. We can revisit this argument by giving here one explicit way to produce fields with this spectrum covariance matrix : one samples first an amplitude $A_0 = b^2$ from any PDF with $\av{A_0} = 1$, $\av{A_0^2}- \av{A_0}^2 = \atot^2$. One then generates independently a Gaussian field $\delta$ with spectrum $P(k)$. The field $b \delta$ has then spectrum $P(k)$ with covariance matrix
\beq
\Cov_{ij} = (1 + \atot^2)\delta_{ij}\frac{2P^2(k_i)}{N_{k_i}} + \atot^2P(k_i)P(k_j).
\enq
Since $\atot^2 \ll 1$ this reduces to form $\eqref{Cov}$ indeed. Note that $b\delta$ has random phases but is not a Gaussian random field due to the fluctuations in $b$. The physical motivation and usefulness of this specific algorithm seem however fairly limited, any field realization looking like a perfect Gaussian map and not like the matter distribution. It may give some physical insights and be of practical use to find a similarly simple but better motivated prescription. 
\newline
\indent
Given the successes of this simple covariance matrix model, it is legitimate to ask whether the technique can be extended deeper into the nonlinear regime. In that regard, we note that the SS coefficient has a straightforward extension within TH13 spectrum response formalism.The trispectrum consistency condition asserted in TH13 suggests that the SS term can be generalized to a function of scale, $\assm(k)$, through $\assm(k) = \sigma_V \partial \ln P(k)/ \partial \delta_b.$
There the stochasticity of the response $\partial_{\delta_b} \ln P(k)$ to the background mode is neglected, and $\delta_b$ is treated as an additional parameter. The covariance matrix remains invertible with the SM formula (or, in the presence of the IS term, with its extension the Woodburry formula \citep{PressEtal07}), recovering a similar result \eqref{FNG}  where essentially our log-amplitude parameter becomes now $\delta_b$, and the denominator is the Gaussian information on $\delta_b$. A decay of the spectrum response, and thus of $\assm(k)$,  on deeply nonlinear scales as seen in \cite{LiEtal2014} would have as a consequence to naturally lift off the SS plateau.  An interesting question is whether an analogous  physical interpretation  and generalization is possible for the IS covariance, that would allow similar analytical insights deep into the nonlinear regime.
\newline
\newline
The authors acknowledge NASA grants NNX12AF83G and NNX10AD53G for support, and thank their reviewer Mark Neyrinck for many useful comments. The research leading to these results has received funding from the European
Research Council under the European Union's Seventh Framework Programme
(FP/2007-2013) / ERC Grant Agreement No. [616170].


\bibliographystyle{mn2e}
\bibliography{bib}

\end{document}